\documentclass[11pt,a4paper]{article}
\usepackage{jinstpub}
\usepackage{mathrsfs}
\usepackage{bm}
\usepackage[separate-uncertainty=true,multi-part-units=single]{siunitx}
\usepackage[version=4]{mhchem}
\usepackage{booktabs}
\usepackage{subfig}
\usepackage{upgreek}

\begin{document}
\title{An exploratory study of a tellurium-loaded liquid scintillator based on water and p-dioxane}
\author[a,b,c]{Ye Liang,}
\author[a,b,c]{Haozhe Sun,}
\author[a,b,c,*]{and Zhe Wang\note[*]{~Corresponding author.}}

\affiliation[a]{Department of Engineering Physics, Tsinghua University, Beijing 100084, China}
\affiliation[b]{Center for High Energy Physics, Tsinghua University, Beijing 100084, China}
\affiliation[c]{Key Laboratory of Particle \& Radiation Imaging (Tsinghua University), Ministry of Education, Beijing 100084, China}

\emailAdd{wangzhe-hep@tsinghua.edu.cn}

\abstract{
Tellurium-loaded liquid scintillators are critical for neutrinoless double-beta decay experiments.
However, conventional organic scintillators are constrained by the limited solubility of organic tellurium compounds compared with that of inorganic ones in water, whereas water-based scintillators are likely constrained by the destabilization of surfactants caused by inorganic tellurium compounds.
In this work, a surfactant-free water-containing route is explored, in which an aqueous telluric acid solution is introduced into a water-miscible organic scintillator comprising p-dioxane, naphthalene, and PPO.
The phase behavior of this system is mapped to delineate homogeneous-mixture domains and to estimate practical upper bounds on tellurium loading.
Optical properties are characterized by UV–visible absorption spectroscopy and fluorescence spectroscopy.
The scintillation light yield is obtained with a relative method that compares to a reference LAB–PPO scintillator.
The measurements demonstrate scintillation quenching induced by water and by telluric acid.
These results provide benchmarks for water-containing and surfactant-free formulations and support the development of high-loading liquid scintillators for future detector design.
}

\keywords{Scintillators, scintillation and light emission processes (solid, gas and liquid scintillators); Double-beta decay detectors; Detector design and construction technologies and materials}

\maketitle

\section{Introduction}

Tellurium-loaded liquid scintillators play an important role in neutrinoless double-beta decay ($0\upnu\upbeta\upbeta$) searches with large-scale detectors. Conventional approaches, primarily developed by collaborations such as SNO+ \cite{thesno+collaboration2021sno}, focus on incorporating tellurium into organic scintillators through the synthesis of tellurium-containing compounds to improve solubility \cite{auty2023method,ding2023novel,suslov2022development,ding2025novel}. The most common strategy involves the dehydration condensation of telluric acid (TeA) with diols to form viscous oligomers. However, these oligomers often exhibit limited solubility in linear alkylbenzene (LAB), necessitating additional solubilizers and ultimately constraining the achievable tellurium loading \cite{auty2023method,ding2023novel}.

Alternatively, tellurium compounds such as telluric acid and tellurite possess high solubility in water, suggesting the potential of water-based scintillators. A major challenge in these systems is the incorporation of sufficient fluor to enable efficient conversion of high-energy particle energy into visible photons detectable by PMTs. Currently, the light yield of water-based liquid scintillators reaches at most about 10\% of that of conventional organic scintillators \cite{caravaca2020characterization,choi2022development,steiger2024development}. Surfactant-assisted methods to load tellurium have been investigated \cite{biller2015sno,shimizu2019double}, but they were not adopted in the final detector designs, such as that of SNO+. Previous studies of emulsion counting have shown that introducing large amounts of water and inorganic salts into surfactant-based systems can lead to phase separation or gelation \cite{zarybnicky1979tritium,bagan2009plastic,bergeron2017phase}, which might be a major concern.

A historical surfactant-free approach based on p-dioxane and naphthalene was reported as early as the 1950s \cite{farmer1952determination,birks1964theorya} and later became known as Bray’s cocktail \cite{bray1960simple}. This formulation was once widely used as a liquid scintillator because of its ability to incorporate aqueous samples while maintaining efficient scintillation. In this system, p-dioxane not only acts as a scintillation solvent capable of dissolving sufficient amounts of PPO, but also exhibits good miscibility with water and compatibility with electrolyte solutions. Naphthalene mitigates quenching effects caused by water and inorganic species, thereby improving the efficiency of energy transfer between solvent and solute \cite{kallmann1956scintillating}. However, Bray’s cocktail is now seldom employed, because p-dioxane tends to form explosive peroxides during storage and also has toxicity, while naphthalene is highly volatile. These drawbacks make the system unsuitable for large-scale neutrino detectors. Nevertheless, its ability to accommodate large fractions of water and inorganic salts while retaining measurable scintillation makes it a useful model system for exploratory studies.

In this work, we explore the performance of surfactant-free, water-containing liquid scintillators using a classic p-dioxane–naphthalene–PPO system capable of incorporating a large quantity of water and inorganic tellurium compounds. We established the ternary phase diagram of the Te–water–p-dioxane system, characterized fluorescence and UV–visible absorption, and measured the scintillation light yield across a range of tellurium concentrations. This work provides an experimental basis for the development of high-loading liquid scintillators and informs the design of future detectors.

\section{Solution preparation}\label{sec:prep}

Telluric acid (TeA, CAS:7803-68-1, purity 99\%, J\&K Scientific Ltd.), p-dioxane (CAS: 123-91-1, purity 99.7\%, stabilized with BHT, water $\leq$ 50~ppm, Shanghai Macklin Biochemical Co., Ltd.), and naphthalene (CAS: 91-20-3, scintillation grade, purity 99\%, Shanghai Macklin Biochemical Co., Ltd.) were used in this study. 2,5-Diphenyloxazole (PPO, CAS:92-71-7, scintillation grade, purity 99\%, Aladdin Scientific Corp.) was employed as the fluor. Linear alkylbenzene (LAB), purchased from Jinling Petrochemical Co., Ltd., was used as the reference scintillator for light-yield measurements. This product was also employed in the Daya Bay~\cite{beriguete2014production} and JUNO~\cite{landini2024distillation} experiments. Ultrapure water was produced using a purification system from Zhongyang Yongkang Environmental Science Co., Ltd.

Solutions of TeA, p-dioxane, water, and naphthalene were prepared according to solubility and phase compatibility constraints. All the containers and cuvettes were rinsed with absolute ethanol, water, and p-dioxane consecutively, and were dried beforehand. Solid TeA was first dissolved in water and the solution was filtered through a \qty{0.22}{\um} membrane to remove particulates, yielding a TeA stock solution. For the preparation of the Te-loaded liquid scintillator, the TeA stock solution was diluted with water to obtain solution~A, while naphthalene and PPO were completely dissolved in p-dioxane to obtain solution~B. Solutions~A and B were then combined and stirred until a clear mixture was obtained. Finally, the container was sealed with a cap and Parafilm\textsuperscript{\textregistered} to minimize naphthalene sublimation. The detailed phase behavior and solubility limits are discussed in the next section.

\section{Solubility phase diagram}\label{sec:phase}

Scintillators for $0\upnu\upbeta\upbeta$ searches require both a high tellurium loading and a large fraction of organic solvent. High tellurium content increases sensitivity, whereas a higher organic fraction generally enhances light yield and thus energy resolution. However, these requirements are mutually constraining: increasing the water fraction improves TeA solubility but typically suppresses scintillation performance. Although p-dioxane is fully miscible with water, the presence of water limits the solubility of TeA. To map this incompatibility, we performed a systematic composition scan of TeA aqueous solutions and p-dioxane mixtures to delineate the phase boundary and identify conditions under which stable, homogeneous solutions can be formed.

To construct the ternary solubility phase diagram, a TeA stock solution (22.2\% by mass, containing 12.3\% Te) was prepared. This stock was then diluted with ultrapure water at selected ratios to produce clear initial solutions. A known amount of p-dioxane was added, followed by stirring for five minutes and settling for another five minutes. Solutions were monitored for the appearance of precipitates. Special attention was paid near the precipitation boundary, where locally formed precipitates could redissolve by diffusion. Each composition was classified as either soluble or insoluble based on the presence or absence of visible precipitation. By repeatedly adding p-dioxane in increments, we generated a series of compositions along a line from the water--TeA axis toward the p-dioxane apex of the ternary diagram. Repeating this procedure for multiple initial water--TeA mixtures enabled an efficient and systematic scan of the phase space.

\begin{figure}[htb]
  \begin{center}
    \includegraphics[width=0.78\textwidth]{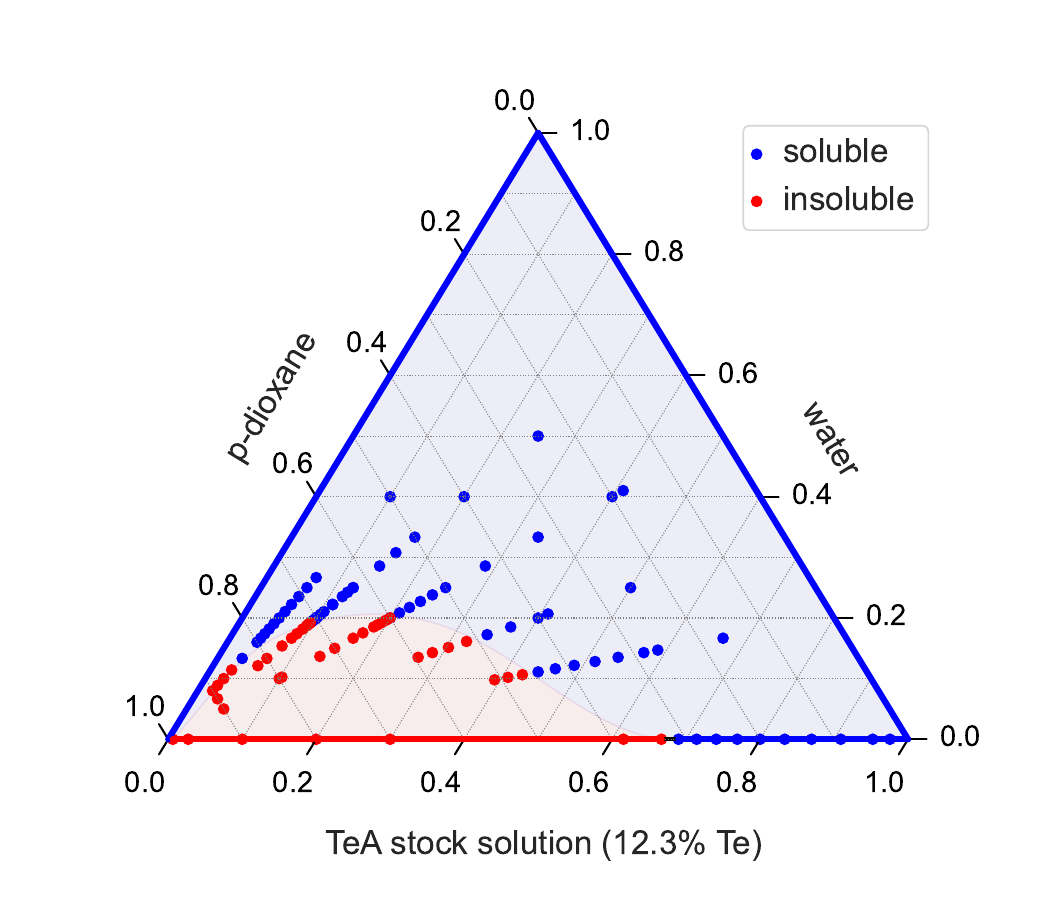}
    \caption{Ternary solubility phase diagram for the TeA--water--p-dioxane system, showing the boundary between soluble and insoluble regions.}\label{fig:phasediagram}
  \end{center}
\end{figure}

The resulting phase diagram is shown in Figure~\ref{fig:phasediagram}. A clear boundary separates the soluble and insoluble regions. For scintillator formulation, compositions should be selected within the soluble region close to this boundary, which allows maximizing the organic solvent content while maintaining the desired tellurium loading. Unless otherwise noted, loadings are reported as wt\%~Te; for reference, $w_{\ce{Te}} \approx 0.556\,w_{\ce{TeA}}$ for TeA (\ce{Te(OH)6}).

\begin{figure}[htb]
\centering
\includegraphics[width=0.68\textwidth]{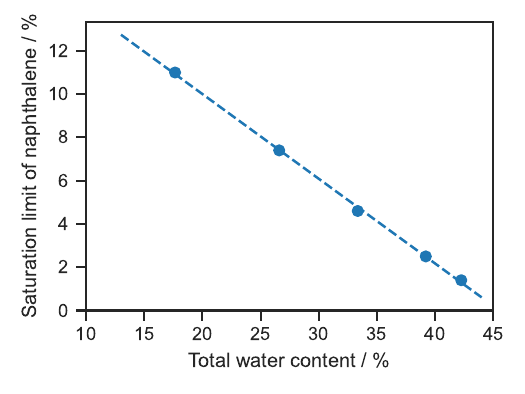}
\caption{Maximum naphthalene solubility as a function of water content in p-dioxane solutions. The dashed line serves as a visual guide.}
\label{fig:naphlim}
\end{figure}

After determining suitable ratios of TeA, water, and p-dioxane, naphthalene was introduced by partially substituting p-dioxane while keeping the total organic fraction invariant. According to experimental observations, naphthalene at low concentrations in p-dioxane did not cause TeA to precipitate, so its effect on the phase boundary was ignored. The upper limit of naphthalene concentration was constrained by the water content in the p-dioxane phase. The maximum soluble naphthalene content as a function of water fraction was determined experimentally (Figure~\ref{fig:naphlim}) and used to guide the choice of naphthalene concentration in different formulations.

\section{Fluorescence spectra}\label{sec:fluo}
\begin{figure}[htb]
  \centering
  \includegraphics[width=0.54\textwidth]{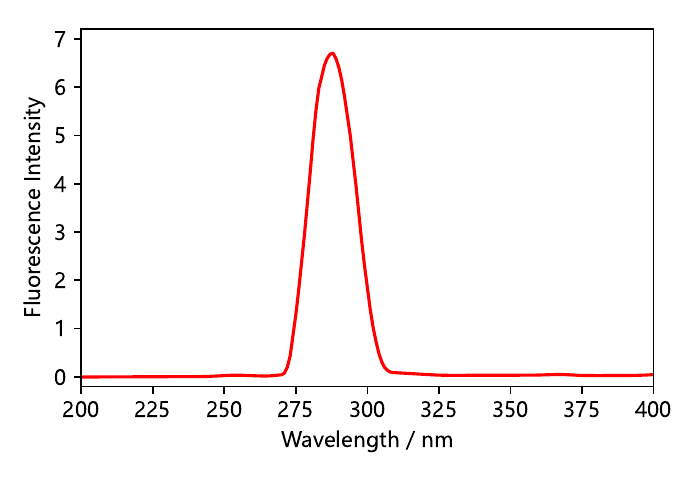}
  \caption{Fluorescence emission spectrum of p-dioxane excited at \qty{185}{nm} with a low-pressure mercury lamp.}
  \label{fig:dioxane_f}
\end{figure}

Fluorescence emission spectra were measured using a Prolab F97Pro fluorescence spectrophotometer (Lengguang Technology Co., Ltd.) with a standard four-pass quartz cuvette of 1~cm path length, transparent in the ultraviolet region. p-Dioxane is effectively excited by \qty{185}{nm} radiation but shows little sensitivity above \qty{200}{nm} \cite{hirayama1970fluorescence}. Therefore, a low-pressure mercury lamp emitting at \qty{185}{nm} was used to excite pure p-dioxane. The resulting fluorescence spectrum is shown in Figure~\ref{fig:dioxane_f}, with an emission peak at \qty{287}{nm} corresponding to the primary radiative transition of p-dioxane.

\begin{figure}[htb]
  \centering
  \subfloat[]{\includegraphics[width=0.54\textwidth]{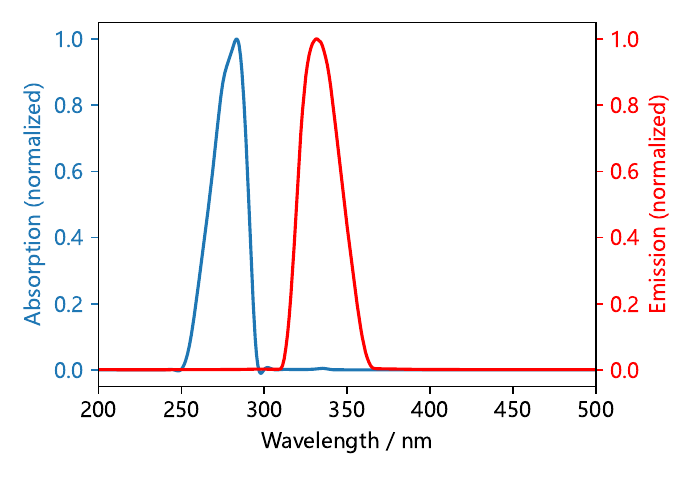}\label{fig:naph_f}}\\
  \subfloat[]{\includegraphics[width=0.54\textwidth]{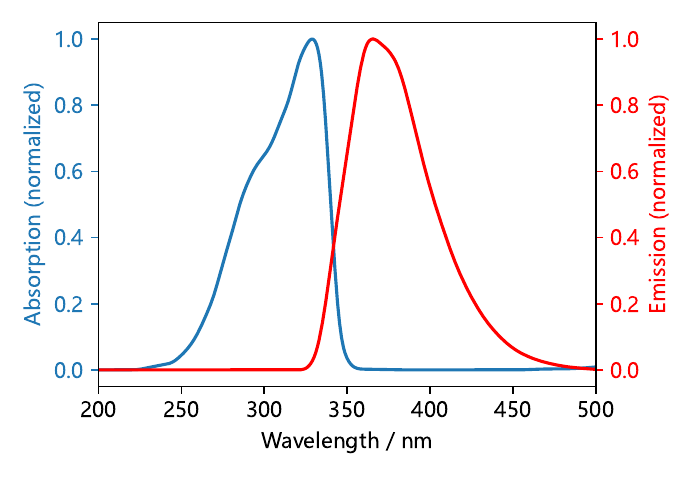}\label{fig:ppo_f}}
  \caption{Fluorescence emission and absorption spectra of (a) \qty{10}{mg/L} naphthalene in ethanol, excited at \qty{283}{nm}, and (b) \qty{10}{mg/L} PPO in ethanol, excited at \qty{330}{nm}.}
\end{figure}

For naphthalene and PPO, ethanol solutions with a concentration of \qty{10}{mg/L} were prepared, and both fluorescence emission and absorption spectra were recorded using a deuterium–xenon light source with excitation wavelength scanning. The spectra are presented in Figures~\ref{fig:naph_f} and \ref{fig:ppo_f}. Naphthalene exhibits a strong absorption peak at \qty{283}{nm} and an emission peak at \qty{330}{nm}, showing a large Stokes shift that minimizes overlap between absorption and emission. PPO shows maximum absorption at \qty{330}{nm} and a corresponding emission peak at \qty{365}{nm}. These peak positions, Stokes shifts, and overall spectral shapes agree well with the literature \cite{berlman1965handbook}.

\begin{figure}[htb]
  \begin{center}
    \includegraphics[width=0.78\textwidth]{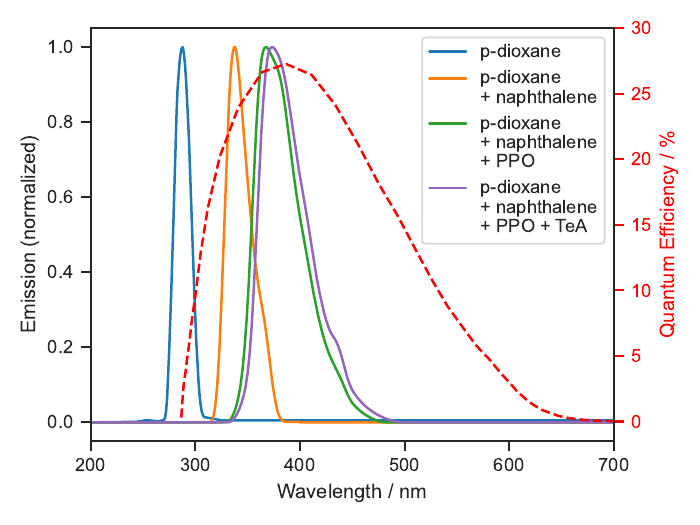}
    \caption{Fluorescence emission spectra of p-dioxane alone and in mixtures with naphthalene and/or PPO, compared with the quantum efficiency of bialkali PMTs (plotted in red).}\label{fig:redshift}
  \end{center}
\end{figure}

Fluorescence spectra under \qty{185}{nm} excitation were also measured for p-dioxane in combination with PPO, naphthalene, or both. The results, together with the quantum efficiency window of the bialkali PMTs \cite{hamamatsuphotonicspmts}, are shown in Figure \ref{fig:redshift}. The progressive redshift of the emission spectrum upon the addition of naphthalene and PPO illustrates the role of each component. While both PPO and naphthalene contribute to enhancing the scintillation light yield, the final emission spectrum corresponds to that of PPO, indicating that PPO acts as the terminal fluor. Naphthalene functions as an intermediate wavelength shifter, facilitating efficient energy transfer from p-dioxane to PPO. This cascading energy transfer mechanism improves overall energy conversion efficiency within the scintillator system.

In addition, the fluorescence emission spectrum of the tellurium-loaded scintillator was measured, as also shown in Figure~\ref{fig:redshift}. A slight redshift of the emission peak was observed upon the introduction of tellurium, suggesting minor changes in the local electronic environment in the scintillation medium.

\section{Absorption spectra}\label{sec:abs}
Ultraviolet–visible attenuation spectra were recorded with a Prolab UV1910 UV–Vis spectrophotometer (Lengguang Technology Co., Ltd.) using a 10~cm quartz cuvette. All liquid samples were measured at room temperature.

\begin{figure}[htb]
\centering
\includegraphics[width=0.78\textwidth]{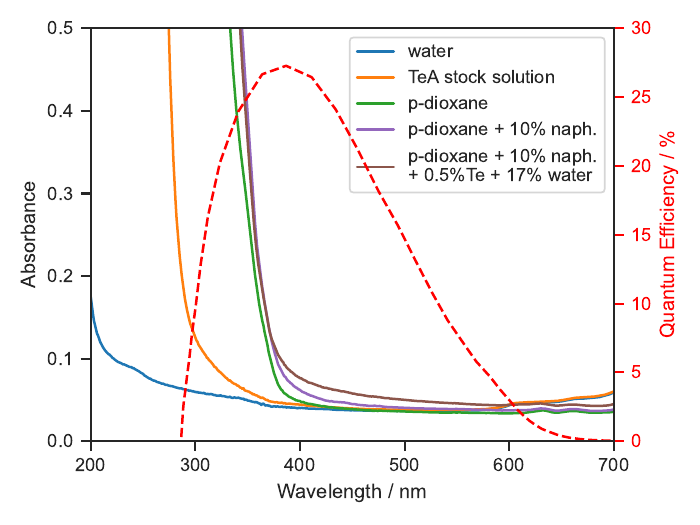}
  \caption{UV-Vis absorption spectrum of the TeA stock solution (12.3\% Te), pure p-dioxane, p-dioxane with 10\% naphthalene, and a 0.5\% Te-loaded p-dioxane scintillator without PPO. The quantum efficiency of bialkali PMTs~\cite{hamamatsuphotonicspmts} is plotted in red for reference.}
\label{fig:TeA_abs}
\end{figure}

The spectra of the following samples were measured and are plotted in Figure~\ref{fig:TeA_abs}:  
(i) the TeA stock solution (12.3\% Te),  
(ii) pure p-dioxane,  
(iii) p-dioxane containing 10\% naphthalene, and  
(iv) a Te-loaded liquid scintillator consisting of p-dioxane with 10\% naphthalene, 0.5\% Te and 17\% water.  
Ultrapure water was also measured as a reference.  
The TeA stock solution exhibits negligible absorption above \qty{360}{nm}, while p-dioxane shows no measurable absorption beyond \qty{400}{nm}.  
Introducing 10\% naphthalene increases the absorbance of p-dioxane by about 0.01 at \qty{400}{nm}, which is caused by the absorption of naphthalene.  
The Te-loaded mixture, which was passed twice through a \qty{0.22}{\um} membrane filter before measurement, displays an additional step of roughly 0.01 in absorbance above \qty{400}{nm} compared with the naphthalene--p-dioxane solution.  
This small offset is likely attributed to light scattering within the liquid. PPO self-absorption can be mitigated by a secondary wavelength shifter~\cite{birks1964theorya} and is not discussed here.

\section{Scintillation light yield}\label{sec:ly}

The relative light yield was measured by comparing the scintillation response of each sample, irradiated with a $\upbeta$ source, to that of a reference scintillator of known yield under identical experimental conditions. The setup comprised a single PMT, a quartz cuvette, and a flash analog-to-digital converter (FADC). Waveforms from $\upbeta$ events were integrated in time to obtain the waveform integrals (in \si{mV\cdot ns}), which are proportional to the charge collected by the PMT up to a constant resistance factor. These integrals were then compared with those of the reference scintillator to extract the relative light yield.

\begin{figure}[htb]
\centering
\includegraphics[width=0.48\textwidth]{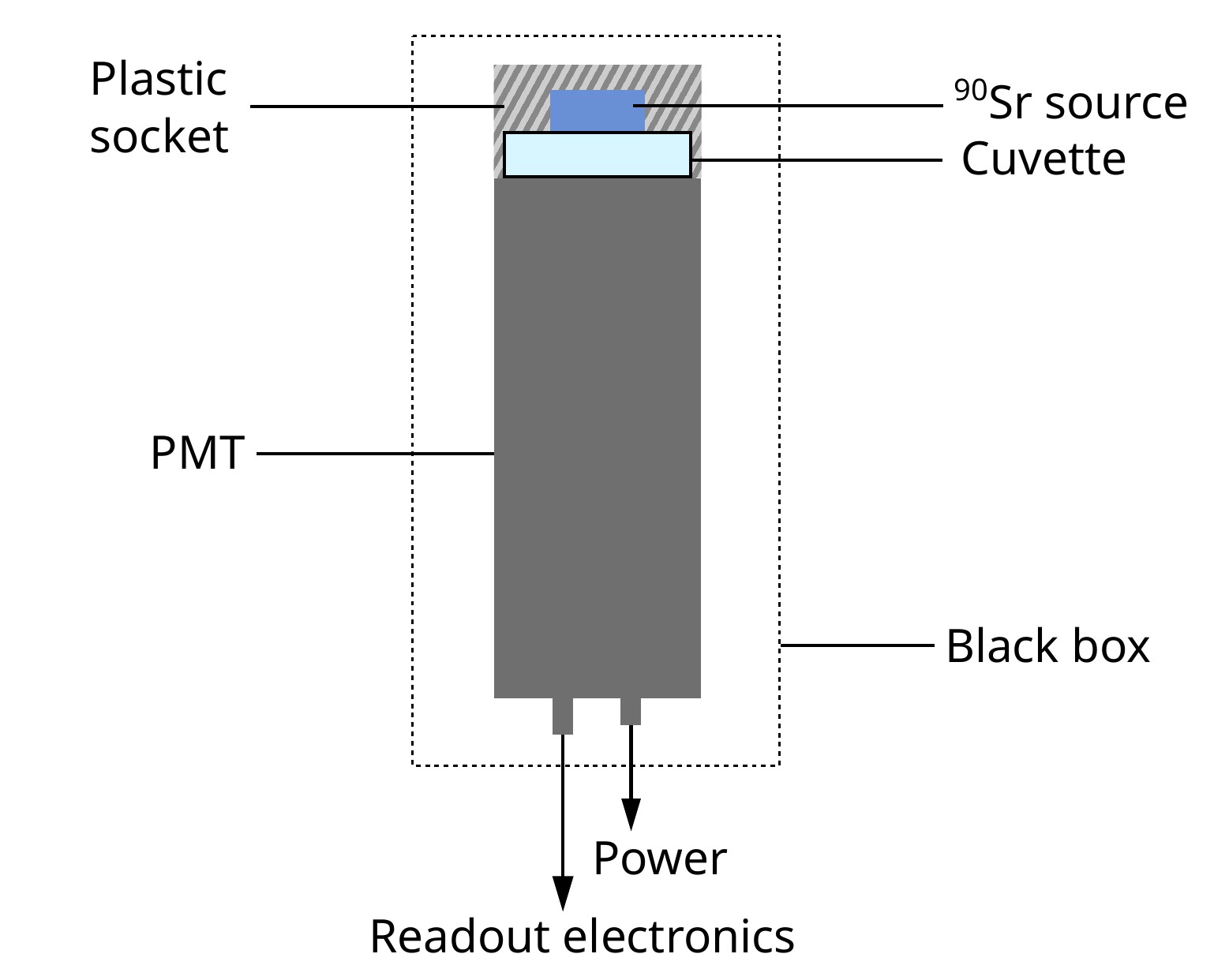}
\caption{Light yield measurement device.}
\label{fig:scintillation_spectrometer}
\end{figure}

The measurement setup is shown in Figure \ref{fig:scintillation_spectrometer}.
A \ce{^{90}Sr}/\ce{^{90}Y} $\upbeta$ source irradiated the liquid sample contained in a 4~cm $\times$ 4~cm $\times$ 1~cm quartz cuvette.
Scintillation light was collected by a 2-inch Hamamatsu R1828-01 bialkali PMT operated at –1500 V inside a light-tight enclosure at room temperature.
The PMT signals were digitized by a CAEN DT5751 flash analog-to-digital converter (FADC) \cite{caen}, which operated in self-trigger mode with a threshold 10 mV below the baseline.
For each sample the data-taking time was kept identical and more than $10^5$ waveforms were recorded.
After each power-up of the PMT and FADC, data taking began only after roughly 10 minutes to allow the system to reach stable operation, and the recorded waveforms were then transferred to a computer for analysis.

\begin{figure}[htb]
  \begin{center}
    \includegraphics[width=0.68\textwidth]{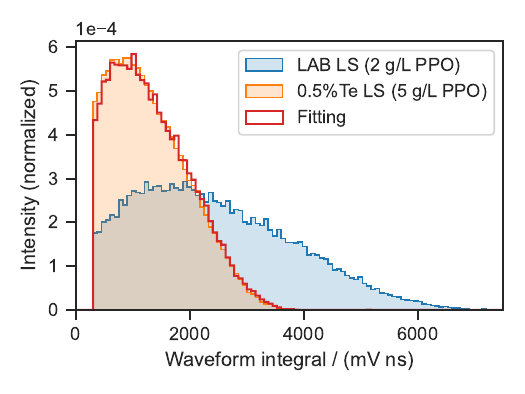}
    \caption{Comparison of waveform integral distributions for a tellurium-loaded scintillator formulation and a reference scintillator (LAB + \qty{2}{g/L} PPO)}\label{fig:fitting}
  \end{center}
\end{figure}

The integral value of an individual waveform is proportional to the charge collected by the PMT, and therefore its distribution reflects the light yield of the liquid scintillator. Figure \ref{fig:fitting} shows the waveform integral distributions for a tellurium-loaded scintillator sample and a reference scintillator. By inferring and comparing the endpoint values of these distributions, the relative light yield can be determined.

The FADC waveform length was 1029~ns with the trigger positioned at 30\% of the record window. The average of the first 150~ns was taken as the baseline. After baseline subtraction, the waveform was integrated from 50~ns before the trigger to 250~ns after the trigger, which was chosen to fully cover the scintillation pulse while leaving pre-trigger margin. The resulting waveform integrals are proportional to the charge collected by the PMT up to a constant resistance factor. Each $\upbeta$ event yields one waveform integral; these values together form the distribution used for the light-yield analysis.

The relative light yield was obtained by comparing the waveform-integral distribution of each sample with that of a reference LAB scintillator consisting of LAB with \qty{2}{g/L} PPO.  
Assuming that the $\upbeta$ source deposits the same energy spectrum in all liquids and that the optical geometry and PMT gain remain unchanged, the two distributions are related by
\begin{equation}
g(X)=k\,f(kX),
\end{equation}
where $f(X)$ is the LAB distribution and $1/k$ gives the relative light yield.  
To determine $k$, the LAB waveform integrals were scaled by trial factors $k$ and rebinned to match the Te-loaded sample.  
The chi–squared statistic
\begin{equation}
\chi^{2}=\sum_i \frac{\left(n_i-\tfrac{N}{M}m_i\right)^2}{n_i+\left(\tfrac{N}{M}\right)^2 m_i}
\end{equation}
was then calculated after events with integrals below \qty{500}{mV\cdot ns} were removed to suppress low-end noise.  
Here $n_i$ and $m_i$ are the $i$-th bin counts of the Te-loaded and scaled LAB histograms, and $N$ and $M$ are their totals.  
The best-fit value of $k$ is obtained by minimizing $\chi^{2}$, and its $1\sigma$ statistical uncertainty, derived from the standard $\Delta\chi^{2}=1$ criterion, is generally less than 1\%.  
Systematic uncertainties arise mainly from solution preparation and contributes about 4.5\%.
Additional contributions come from refractive-index–related geometry effects; these amount to about 2.0\% and were estimated with a Monte-Carlo ray-tracing simulation.
A further source is possible PMT gain drift, which contributes about 0.3\% as determined from single-photon checks.
Combining the statistical and systematic components gives an overall uncertainty in relative light-yield of about 5\%. The formulations and measurement results for all samples are summarized in Table \ref{tab:points}.

\begin{table}[htb]
\begin{center}
  \caption{Tellurium-free and Tellurium-loaded p-dioxane liquid scintillator formulations and relative light yields, normalized to LAB + \qty{2}{g/L} PPO. The percentages are by mass.}\label{tab:points}
\begin{tabular}{@{}cccccccc@{}}
\toprule
  \# & Te / \% & TeA+water / \% & p-Dioxane / \% & Naphthalene / \% & PPO / (g/L) & Relative light yield  \\ \midrule
1         & 0       & 0          & 100            & 0                & 5           & 0.555 $\pm$ 0.028 \\
2         & 0       & 10         & 90             & 0                & 5           & 0.294 $\pm$ 0.015 \\
3         & 0       & 20         & 80             & 0                & 5           & 0.213 $\pm$ 0.011 \\
4         & 0       & 30         & 70             & 0                & 5           & 0.167 $\pm$ 0.008 \\
5         & 0       & 40         & 60             & 0                & 5           & 0.141 $\pm$ 0.007 \\
6         & 0       & 50         & 50             & 0                & 5           & 0.121 $\pm$ 0.006 \\
7         & 0       & 0          & 95             & 5                & 5           & 0.913 $\pm$ 0.046 \\
8         & 0       & 10         & 85             & 5                & 5           & 0.667 $\pm$ 0.033 \\
9         & 0       & 20         & 75             & 5                & 5           & 0.489 $\pm$ 0.024 \\
10        & 0       & 25         & 70             & 5                & 5           & 0.433 $\pm$ 0.022 \\
11        & 0       & 0          & 90             & 10               & 5           & 1.003 $\pm$ 0.050 \\
12        & 0       & 10         & 80             & 10               & 5           & 0.770 $\pm$ 0.038 \\
13        & 0       & 15         & 75             & 10               & 5           & 0.661 $\pm$ 0.033 \\
14        & 0.125   & 20         & 75             & 5                & 5           & 0.452 $\pm$ 0.023 \\
15        & 0.250   & 20         & 75             & 5                & 5           & 0.413 $\pm$ 0.021 \\
16        & 0.375   & 20         & 75             & 5                & 5           & 0.381 $\pm$ 0.019 \\
17        & 0.500   & 20         & 75             & 5                & 5           & 0.364 $\pm$ 0.018 \\
18        & 0.625   & 20         & 75             & 5                & 5           & 0.349 $\pm$ 0.017 \\
19        & 0.125   & 15         & 75             & 10               & 5           & 0.623 $\pm$ 0.031 \\
20        & 0.250   & 15         & 75             & 10               & 5           & 0.580 $\pm$ 0.029 \\
21        & 0.375   & 15         & 75             & 10               & 5           & 0.537 $\pm$ 0.027 \\ 
22        & 0.500   & 17.35      & 71.55          & 11.10            & 5           & 0.527 $\pm$ 0.026 \\
23        & 1.000   & 26.60      & 65.90          & 7.50             & 5           & 0.324 $\pm$ 0.016 \\
24        & 1.500   & 33.35      & 62.05          & 4.60             & 5           & 0.215 $\pm$ 0.011 \\
25        & 2.000   & 39.20      & 58.30          & 2.50             & 5           & 0.141 $\pm$ 0.007 \\
26        & 2.500   & 42.25      & 56.35          & 1.40             & 5           & 0.106 $\pm$ 0.005 \\ \bottomrule
\end{tabular}
\end{center}
\end{table}

\begin{figure}[htb]
\centering
\includegraphics[width=0.65\textwidth]{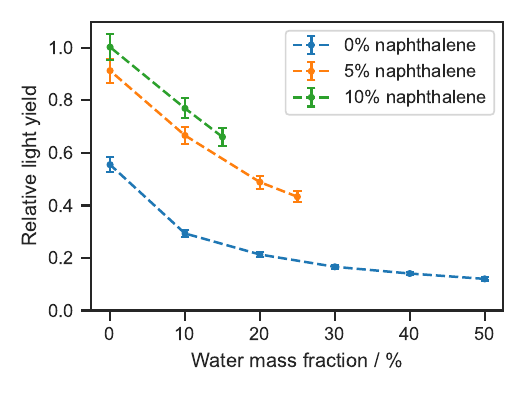}
  \caption{Relative light yield (normalized to LAB + \qty{2}{g/L} PPO) as a function of water content in dioxane-PPO solution, all containing \qty{5}{g/L} PPO.}
\label{fig:water_dioxane}
\end{figure}

The effect of water content on the light yield of dioxane based scintillators containing \qty{5}{g/L} PPO was systematically investigated as shown in Figure \ref{fig:water_dioxane} (corresponds to Table \ref{tab:points} \#1--\#13). 
Water showed a significant quenching effect. Without naphthalene the addition of 10\% water lowered the relative light yield by about 47\% from 0.555 to 0.293.
Adding naphthalene markedly enhanced the light yield and at the same time mitigated this quenching.
With 5\% naphthalene the same 10\% water addition reduced the yield by about 27\% from 0.913 to 0.667.
With 10\% naphthalene the 10\% water addition reduced the yield by about 23\% from 1.003 to 0.770.
The solubility limit of naphthalene decreases as the water content increases and this limits the maximum achievable naphthalene concentration.
To achieve high light yield the water content should therefore be kept as low as practicable and the naphthalene concentration increased to the solubility limit.

\begin{figure}[htb]
\centering
\includegraphics[width=0.65\textwidth]{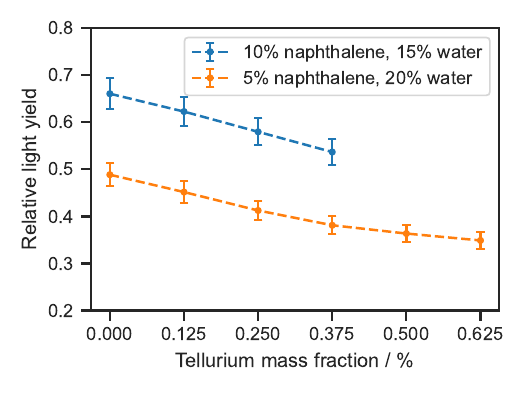}
  \caption{Relative light yield (normalized to LAB + \qty{2}{g/L} PPO) as a function of tellurium content in water-dioxane-PPO solution, all containing \qty{5}{g/L} PPO.}
\label{fig:water_dioxane_Te}
\end{figure}

The effect of telluric acid on the light yield was measured for two series of p-dioxane scintillators containing \qty{5}{g/L} PPO. One series contained 5\% naphthalene and 20\% water, and the other contained 10\% naphthalene and 15\% water. In each series, the tellurium concentration was varied from 0 up to the respective solubility limits. The resulting relative light yields are shown in Figure~\ref{fig:water_dioxane_Te} (corresponding to Table \ref{tab:points} \#9, \#13--\#21). For the 5\% naphthalene and 20\% water series, the yield decreased from 0.489 at 0\% Te to 0.349 at 0.625\% Te, a reduction of about 29\%. For the 10\% naphthalene and 15\% water series, the yield decreased from 0.661 at 0\% Te to 0.537 at 0.375\% Te, a reduction of about 19\%. These two trends indicate that telluric acid alone causes a reproducible loss in light yield. As shown in Figure~\ref{fig:TeA_abs}, the aqueous tellurium stock solution exhibits negligible absorption in the PMT-sensitive wavelength range. This observation suggests that telluric acid may interfere with nonradiative energy-transfer processes in the scintillator.

\begin{figure}[!htbp]
  \begin{center}
    \includegraphics[width=0.68\textwidth]{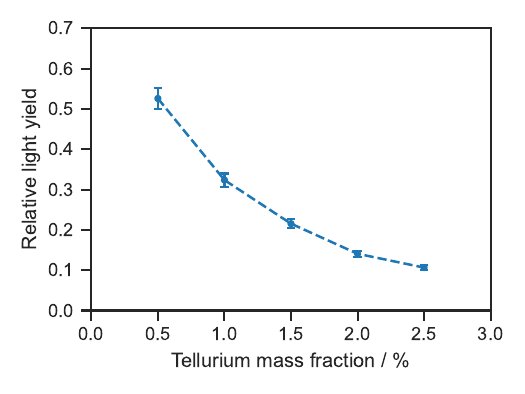}
    \caption{Relative light yield (normalized to LAB + \qty{2}{g/L} PPO) for Te-doped p-dioxane-based scintillators prepared with minimal water and maximal naphthalene allowed by the phase diagram, all containing \qty{5}{g/L} PPO.}\label{fig:RLY}
  \end{center}
\end{figure}

Several scintillator formulations with different tellurium contents were finally prepared using the minimum amount of water and the maximum amount of naphthalene within the limits of the phase diagram (Table \ref{tab:points} \#22--\#26). All samples contained \qty{5}{g/L} PPO. Their relative light yields are presented in Figure \ref{fig:RLY}.

\section{Summary}\label{sec:sum}
This work presents an exploratory study of a tellurium-loaded liquid scintillator based on water and p-dioxane. The solubility limits, optical properties, and scintillation light yield of the representative system are characterized. A clear trade-off was observed between tellurium loading and the organic phase. When the organic phase comprised 80\% by mass of the mixture, the maximum tellurium content was about 0.5\% Te. When the organic phase fraction was reduced to 60\%, the maximum tellurium content increased to about 2.5\%. At this composition, the best formulation achieved only about 20\% of the light yield of the 0.5\% Te scintillator. The cause of the reduction was investigated. Besides water, telluric acid alone was observed to cause the loss in light yield. Since UV–visible spectra showed no appreciable absorption in the PMT-sensitive region by the aqueous tellurium solution, the loss associated with the tellurium compound implied a interference with the nonradiative energy-transfer channels in the scintillator. Further study is needed to reveal the mechanism by which inorganic dopants influence the scintillation light yield, which may help improve the performance of both organic and water-based scintillators. This study characterizes a typical surfactant-free and water-containing scintillator. The results provide practical guidance for selecting high-loading formulations and inform design options for future detectors.

\section{Acknowledgement}
This work is supported in part by the National Natural Science Foundation of China (No.~12141503), the Ministry of Science and Technology of China (No.~2022YFA1604704), and the Key Laboratory of Particle \& Radiation Imaging (Tsinghua University).

\bibliographystyle{JHEP} 
\bibliography{Library.bib}
\end{document}